\documentclass[11pt,twoside]{article}
\usepackage{asp2010}

\resetcounters

\begin{document}

\title{Aspects of Multi-Dimensional Modelling of Substellar Atmospheres}
\author{Ch.~Helling$^{1}$, E.~Pedretti$^{1}$, S.~Berdyugina$^{2}$, A.A.~Vidotto$^1$,  B.~Beeck$^{3,4}$, E.~Baron$^{5,6}$, A.P. Showman$^7$,  E.~Agol$^8$, D. Homeier$^4$
\affil{$^1$  SUPA, School of Physics \& Astronomy, University St Andrews, St Andrews, North Haugh, KY16 9SS, UK}
\affil{$^2$ Kiepenheuer Institut f\"ur Sonnenphysik, Sch\"oneckstr. 6, D-79104 Freiburg, Germany}
\affil{$^3$ Max Planck Institute for Solar-System Research, Max-Planck-Stra\ss e 2, 37191 Katlenburg-Lindau, Germany}
\affil{$^4$ Georg August University, Institute for Astrophysics, Friedrich-Hund-Platz 1, 37077 G\"ottingen, Germany}
\affil{$^5$Homer L.~Dodge Department of Physics and Astronomy,
  University of Oklahoma, 440 W Brooks, Rm 100, Norman, OK 73019 USA} 
\affil{$^6$Computational Research Division, Lawrence Berkeley
        National Laboratory, MS 50F-1650, 1 Cyclotron Rd, Berkeley, CA
        94720 USA}}
\affil{$^7$ Department of Planetary Sciences,
University of Arizona, Tucson, AZ, USA}
\affil{$^8$Dept. of Astronomy, Box 351580, University of Washington, Seattle, WA 98195 }

\begin{abstract}
Theoretical arguments and observations suggest that the atmospheres of
Brown Dwarfs and planets are very dynamic on chemical and on physical
time scales. The modelling of such substellar atmospheres has, hence,
been much more demanding than initially anticipated. This
Splinter\footnote{\small
http://star-www.st-and.ac.uk/$\sim$ch80/CS16/MultiDSplinter$_{-}$CS16.html}
has combined new developments in atmosphere modelling, with novel
observational techniques, and new challenges arising from planetary
and  space weather observations.
\end{abstract}

\section{Introduction}

A rich molecular gas-phase chemistry coupled with cloud formation
processes determines the atmosphere spectra of very low-mass, cool
objects. Interferometry (E. Pedretti, Sect.~\ref{s:pedretti}) and
polarimetry (S. Berdyugina, Sect.~\ref{s:berdyugina}) can potentially
provide more insight. However, present day interferometers are not
capable of surface imaging Brown Dwarfs and planets due to financial
constraints. Polarimetry, as a novel planet detection method, benefits
from Rayleigh scattering on high-altitude sub-$\mu$m cloud
particles. Such clouds were predicted to form by non-equilibrium
processes several years ago (Woitke \& Helling 2004). Wavelength
dependent transit timing may reveal the interaction of the planetary
exosphere with the stellar corona, and hence, limits may be set
on planetary magnetic field strengths (A.A. Vidotto,
Sect.~\ref{s:vidotto}). Radiative MHD simulations suggest that
magnetic field driven convection significantly changes in fully
convective objects compared to the Sun: M-dwarfs are suggested to
exhibit darker magnetic structures (B. Beeck,
Sect.~\ref{s:beeck}). Studies of multi-dimensional radiative transfer
emphasize that full solutions of physical problems are needed to access
limits  approximations (E.~Baron,
Sect.~\ref{s:baron}). The superrotation observed in planetary
atmospheres is suggested to result from standing Rossby waves
generated by the thermal forcing of the day-night temperature
difference (A.P. Showman, Sect.~\ref{s:showman}). A search for
transit-time variations at 8 $\mu$m reveals a difference between the
transit and the secondary eclipse timing after subtracting stellar
variability, and hence, confirms the superrotation on HD~189733b
(E. Agol, Sect.~\ref{s:agol}). Results of multi-dimensional
simulations are starting to be used as input for 1D model atmospheres for
synthetic spectra production (D. Homeier, Sect.~\ref{s:homeier}).

\section{Combined interferometry for substellar variability (Ettore Pedretti)}\label{s:pedretti}

\begin{figure}
\includegraphics[scale=.5]{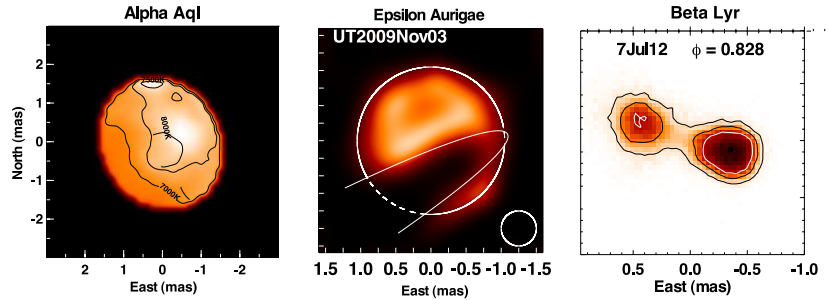}
\centering
\caption{CHARA results. From left: Alpha Aql, first image of a main sequence star other than the Sun; the dust disc orbiting an unseen companion and eclipsing the star $\epsilon$ Aurigae;  $\beta$ Lyr, the closest interacting binary star ever imaged.}
\label{altair}
\end{figure}

Optical and infrared long--baseline interferometry allows high--resolution imaging that is out of reach for the current large telescope facilities and for the planned 30m class telescopes. Examples of typical and future targets for long--baseline interferometry are  stellar surfaces, planet--forming discs, active galactic nuclei and extrasolar planets. The main interferometric facilities in the northern hemisphere are the center for high angular resolution astronomy (CHARA) array and the Keck interferometer. CHARA is a visible and infrared interferometer composed of 6 one--metre telescopes on a 330m maximum baseline (see Pedtretti et al. 2009).
%\cite{2009NewAR..53..353P}. 
The Keck interferometer is composed of two 10m telescopes on a 85m baseline and works mainly in the infrared. The main facility in the southern hemisphere is the very large telescope interferometer (VLTI), composed of four 8m telescopes and four 2m telescopes on a 200m maximum baseline. The Sidney university stellar interferometer (SUSI) in Australia has the longest available baseline in the world (640m) but so far has only used up to 80m baselines. Previous generation interferometers provided unique science by measuring the diameters of stars with two telescopes or by providing simple model dependent imaging combining up to 3 telescopes (Berger et al. 2001, Monnier et al. 2003, Pedretti et al. 2009)
%\citep{2001A&A...376L..31B, 2002SPIE.4838..1127, 
%          2008ApJ...679..746R, 2009MNRAS.397..325P}. 
Model--independent imaging of complex objects was achieved quite recently at the CHARA array, that obtained the first image of a main-sequence star, Altair (Monnier et al. 2007).
%\cite{2007Sci...317..342M}. 
CHARA also imaged the most distant eclipsing system, the star $\beta$~Lirae and witnessed the spectacular eclipse from the $\epsilon$ Aur system 
%\citep{2010Natur.464..870K}
  (Kloppenborg et al. 2010;  Fig.~\ref{altair}). The VLTI imaged the young stellar object IRAS 13481-6124 
 %\citep{2010Natur.466..339K}.
(Kraus et al. 2010).

An interesting question is whether interferometry could resolve brown dwarfs and provide the same sort of high--resolution pictures offered to its larger stellar cousins. $\epsilon$~Indi~B is the nearest brown dwarf %\citep{2003A&A...398L..29S} 
(Scholz et al 2003).
The distance is 3.6 pc, corresponding to an angular diameter of 0.3 milliarcseconds, and a magnitude at H band Mh = 11.3. $\epsilon$~Indi~B is in the southern hemisphere, therefore it is only accessible by the VLTI and SUSI. The VLTI does not have long enough baselines, its maximum baseline  being 200m. SUSI with its 640m baselines would achieve in the infrared, at H band a resolution of 0.5~milliarcseconds, therefore it could measure the diameter and effective temperature of $\epsilon$~Indi~B if its bolometric flux was known. However SUSI has never used baselines longer than 80m and it is not sensitive enough to reach Mh = 11.3, since it uses small 10cm siderostats. Resolved imaging of brown-dwarfs is out of reach for the present interferometric facilities. Brown--dwarfs are at least as challenging to image as Jupiter--size planets. A facility the size of the Atacama large millimetre array (ALMA) at infrared wavelengths would possibly achieve imaging of brown--dwarfs and Jupiter--size planets but it is unlikely that such facility will be financed in the short term. The remaining question is what could existing interferometers do in term of science, other than resolved imaging of the atmosphere of a brown--dwarf. The recent detection of brown dwarfs in binary systems within 5 AU from the main star from Corot (see CoRoT-3b, 
%\citep{2009yCat..34910889D}, 
%(Deleuil et al. 2009),
CoRot-6b and Super-WASP soon) opens up interesting possibilities. Many rejected planets in radial-velocity surveys may be brown dwarfs therefore there may be a large number available targets. Interferometry could potentially characterise brown dwarfs in binary and multiple systems very close to a brighter, more massive companion star through precision closure-phase measurement 
%\citep{2009PhDT.........7Z}.
(Zhao 2009).  Closure-phase nulling 
%\citep{2009A&A...498..321C}, 
(Chelli et al 2009) a special case of precision closure--phase, where measurement are performed around nulls of visibility function and produce large change of closure-phase is potentially more sensitive. Interferometry could yield spectral and flux information about the brown dwarf and derive the mass of the brown--dwarf by measuring the inclination of the orbit in combination with radial velocity measurements.

\section{Polarized Light in stars and planets (Svetlana Berdyugina)}\label{s:berdyugina}

Polarimetry is a powerful technique for revealing hidden structures in
astrophysical objects, far beyond spatial resolution provided by
direct imaging at any telescope. Polarization is a fundamental
%physical 
property of the light.
% and, generally speaking, all light in
%the universe is polarized to some degree, as there are no perfectly
%symmetric astrophysical objects. 
It is its incredible sensitivity to
asymmetries that empowers polarimetry and allows us to look inside
unresolved structures. For instance, light can become polarized when
it is scattered, or passes through magnetized matter, or is absorbed
in an environment illuminated by anisotropic radiation.

Using molecular spectropolarimetry of cool stars enables us to obtain
the first 3D view of starspots with the strongest field and
the coldest plasma on the stellar surface. Such a phenomenon is common
among stars possessing convective envelopes, where magnetic fields are
believed to be generated (Berdyugina 2005).
%\citep{Berdyugina2005lr}. 
However, until recently magnetic fields have never been measured
directly inside starspots. By selecting molecular lines which
preferably form in cool starspots and at different heights in their
atmosphere, such as TiO, CaH, and FeH, starspots
and their internal structure are resolved (Berdyugina
2011).%\citep{Berdyugina2011}.
A new diagnostic technique of embedded stars and inner parts of protoplanetary disks based on radiative pumping of absorbers has been pioneered by (Kuhn et al. 2007).
%\citet{Kuhnetal2007}. 
In the presence of anisotropic incident radiation, e.g. clumps of protoplanetary material illuminated by a star, the lower state magnetic sublevels of atoms or molecules in the intervening gas become unequally populated (even in the absence of magnetic fields). Such an 'optically pumped' gas will result in polarized line absorption along the line of sight. This provides novel insights into the structure and evolution of the innermost parts of circumstellar disks which are inaccessible to any other technique (Kuhn et al. 2011).
%\citep{Kuhnetal2011}. 

\begin{figure}[!ht]
\plottwo{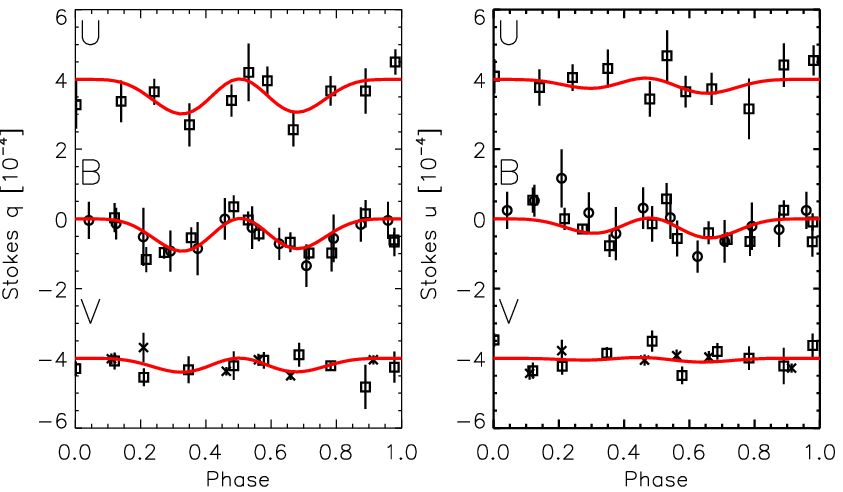}{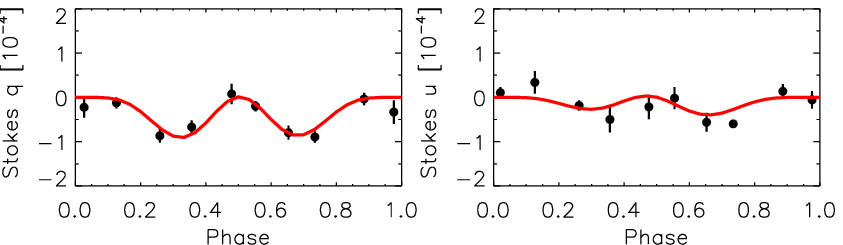}
\caption{Polarimetric data (normalized Stokes $q$ and $u$) for HD189733b in $UBV$ bands. {\it Left:}
squares -- Berdyugina et al. (2011),
%\citet{Berdyuginaetal2011} 
open circles --  binned $B$-band data from Berdyugina et al. (2008),
%\citep{Berdyuginaetal2008} 
%\citet{Wiktorowicz2009} 
crosses --  broad-band filter  data by Wiktorowicz (2009)
centered at the $V$-band. The $U$ and $V$ data are shifted in vertical by $\pm$4$\cdot$10$^{-4}$ for clarity. {\it Right:} All the $U$ and $B$ data from the years 2006-Ð2008 binned together. The mean error of the binned data is 1.7$\cdot$10$^{-5}$. Curves are the best-fit solutions for a model atmosphere with Rayleigh scattering on dust particles. 
%The normalized $\chi^2$ of the fit is 1.16. The standard deviation of the fit is 1.1$\cdot$10$^{-5}$.}
}
\label{fig:obs}
\end{figure}

Detecting planetary atmospheres in polarized light provides a direct probe of exoplanets outside transits. The light scattered in the planetary atmosphere is linearly polarized and, when the planet revolves around the parent star, the polarization varies. Thus, the observed polarization variability exhibits the orbital period of the planet and reveals the inclination, eccentricity, and orientation of the orbit as well as the nature of scattering particles in the planetary atmosphere. HD189733b is a very hot Jupiter and the first exoplanet detected in polarized light  (Berdyugina et al 2008).
%\citep{Berdyuginaetal2008,Berdyuginaetal2011}. 
The observed polarization (Fig.~\ref{fig:obs}) is caused by Rayleigh scattering, possibly on 20\,nm MgSiO$_3$ dust condensates 
(Berdyugina 2011).
%\citep{Berdyugina2011}. 

\section{ Stellar influence on planet atmosphere is shocking (Aline A. Vidotto)}\label{s:vidotto}

WASP-12b is a transiting giant planet that was first identified in an
optical photometric transit survey (Hebb et al. 2009).
%\citep{hebb2009}. 
More recently, transit observations were also done in the near-UV 
(Fossati et al 2010a),
%\citep{fossati2010}, 
revealing that while the time of the egress of the transit occurs almost simultaneously for both the optical and the near-UV observations, the ingress is first seen in the near-UV. This asymmetric behavior in the planet light curve has been explained by the presence of asymmetries in the planetary exosphere.

Motivated by this difference in transit durations, we proposed a model where the interaction of the stellar coronal plasma with the planet is able to modify the structure of the outer atmosphere of WASP-12b 
(Vidotto et al. 2010a; Paper1)
%\citep*{paper1}. 
WASP-12b is a giant planet with $M_p=1.41~M_J$ and $R_p = 1.79~R_J$,
where $M_J$ and $R_J$ are the mass and radius of Jupiter,
respectively. It orbits its host star (a late-F star, with $M_* =
1.35~M_\odot$, $R_*=1.57~R_\odot$) at an orbital radius of $R_{\rm
orb}=0.023~{\rm AU}=3.15~R_*$, with an orbital period of $P_{\rm orb}
= 1.09$~d. Due to such a close proximity to the star, the flux of
coronal particles impacting on the planet comes mainly from the
azimuthal direction, as the planet moves at a relatively high
Keplerian orbital velocity of $u_K = (G M_*/R_{\rm orb})^{1/2}\sim
230$~km~s$^{-1}$ around the star. Therefore, stellar coronal material
is compressed ahead of the planetary orbital motion, possibly forming
a bow shock ahead of the planet.  The condition for the formation of a
bow shock is that the relative motion between the planet and the
stellar corona is supersonic. Although we know the orbital radius of
WASP-12b, we do not know if at this radius the stellar magnetic field
is still capable of confining its hot coronal gas, or if this plasma
escapes in a wind (see Paper 1). In the first case, where the coronal
medium around the planet can be considered in hydrostatic equilibrium,
the velocity of the particles that the planet `sees' is supersonic if
$\Delta u = |u_K-u_\varphi|> c_s$, where $u_\varphi= 2 \pi R_{\rm
orb}/P_*$ is the azimuthal velocity of the stellar corona, $c_s$ is
the sound speed, and $P_*$ is the stellar period of rotation. From
observations of the sky projected stellar rotation velocity,
$P_*\gtrsim 17$~days
%\citep{fossati2010b}. 
(Fossati et al. 2010b). This implies that for a coronal temperature $T
\lesssim (4 - 5) \times 10^6$~K, shock is formed around
WASP-12b. Although stellar flares can raise the coronal plasma
temperature above these values, it is unlikely that a corona would be
hotter than this.

If we take the observationally derived stand-off distance from the
shock to the center of the planet ($\sim 4.2~R_p$, Lai et al. 2010) as
approximately the extent of the planetary magnetosphere $r_M$, we
showed
%\citet{paper1} 
that pressure balance between the coronal total pressure (i.e., ram, thermal, and magnetic pressures) and the planet total pressure requires that
\begin{equation}\label{eq.equilibrium1}
B_c(R_{\rm orb}) \simeq B_p(r_M) ,
\end{equation}
where $B_c(R_{\rm orb})$ is the magnetic field intensity of the star
at $R_{\rm orb}$ and $B_p(r_M)$ is the magnetic field intensity of the
planet at $r_M$. Note that we neglected the ram and thermal pressures
in previous equation. Assuming that both the stellar and the planetary
magnetic fields can be described as dipoles, from
Eq.~(\ref{eq.equilibrium1}), we have
\begin{equation}
B_p = B_* \left(  \frac{R_*/R_{\rm orb}}{R_{p}/r_M}  \right)^3 = B_* \left(  \frac{1/3.15}{1/4.2}  \right)^3\simeq 2.4 B_* ,
\end{equation}
where $B_*$ and $B_p$ are the magnetic field intensities at the stellar and planetary surfaces, respectively. Adopting the upper limit of the stellar magnetic field of $10$~G suggested by 
%\citet{fossati2010b}, 
Fossati et al. (2010b),
our model predicts a maximum planetary magnetic field of about $24$~G. It is likely that shock formation around close-in planets is a common feature of transiting systems. In fact, in a follow-up work Vidotto et al. (2010b), we showed that about $36$ out of $92$ known transiting systems (as of Sept/2010) would lie above a reasonable detection threshold. For these cases, the observation of bow-shocks may be a useful tool in setting limits on planetary magnetic field strengths.

\section{MHD simulations reveal crucial differences between solar and very-cool star magnetic structures (Benjamin Beeck, Manfred~Sch\"ussler, Ansgar~Reiners)}\label{s:beeck}

Cool main-sequence stars of spectral types F through L have a thick
convective envelope or are fully convective. In many of such stars,
magnetic fields of various strengths have been detected.  In the Sun,
the surface magnetic field is observed to be highly structured owing
to its interaction with the convective flows. 
%This has significant
%impact on the magnetic signatures in spectral lines that are used to
%detect and measure the field. 
In contrast to the Sun, the structure
and properties of magnetic fields on cool stars are unknown. In
the absence of spatially resolved observations, the effect of the
magnetic structure on signatures of the magnetic field can be
evaluated by numerical simulations of the magneto-convective
processes.
Using the MHD code MURaM, we carried out 3D radiative
magnetohydrodynamic simulations of the convective and magnetic
structure in the surface layers (uppermost part of the convection zone
and photosphere) of main-sequence stars of spectral types F3 to
M2. 
%The simulation results were analyzed in terms of sizes and
%properties of the convection cells (granules) and magnetic flux
%concentrations as well as velocity, pressure, density, and temperature profiles. 
 The code is a ``box-in-the-star'' code that
solves the equations of (non-ideal) MHD in three spatial dimensions
with constant gravitational acceleration. It includes 
%the relevant physical processes such as 
compressibility, partial ionization, and
non-grey radiative energy transport (for details see V\"ogler et al
2005).
%\cite{MURaM1}.\\
To fit the surface conditions for different stellar spectral types, gravity and effective temperature were adjusted and the opacity bins were recalculated. The size of the computational box and the spatial resolution were modified in order to cover the relevant length scales of the different convection patterns.\\
The model grid comprises six main-sequence stars of spectral types F3, G2, K0, K5, M0, and M2. 
The start models were run with $\mathbf{B}\equiv \mathbf{0}$ until a quasi-stationary state was reached. Then, a homogeneous vertical magnetic field with the field strength $B_0=20\,\mathrm{G}$, $100\,\mathrm{G}$, or $500\,\mathrm{G}$ was introduced.

\begin{figure}
{\ }\\*[-0.0cm]
\centering
\begin{tabular}{cc}
\multicolumn{2}{c}{$I_{\mathrm{bol}}(\mu=1)$ \qquad\qquad G2V \qquad\qquad $B_z(\tau_{\mathrm R}\approx 1)$} \\
\includegraphics[width=3.7cm]{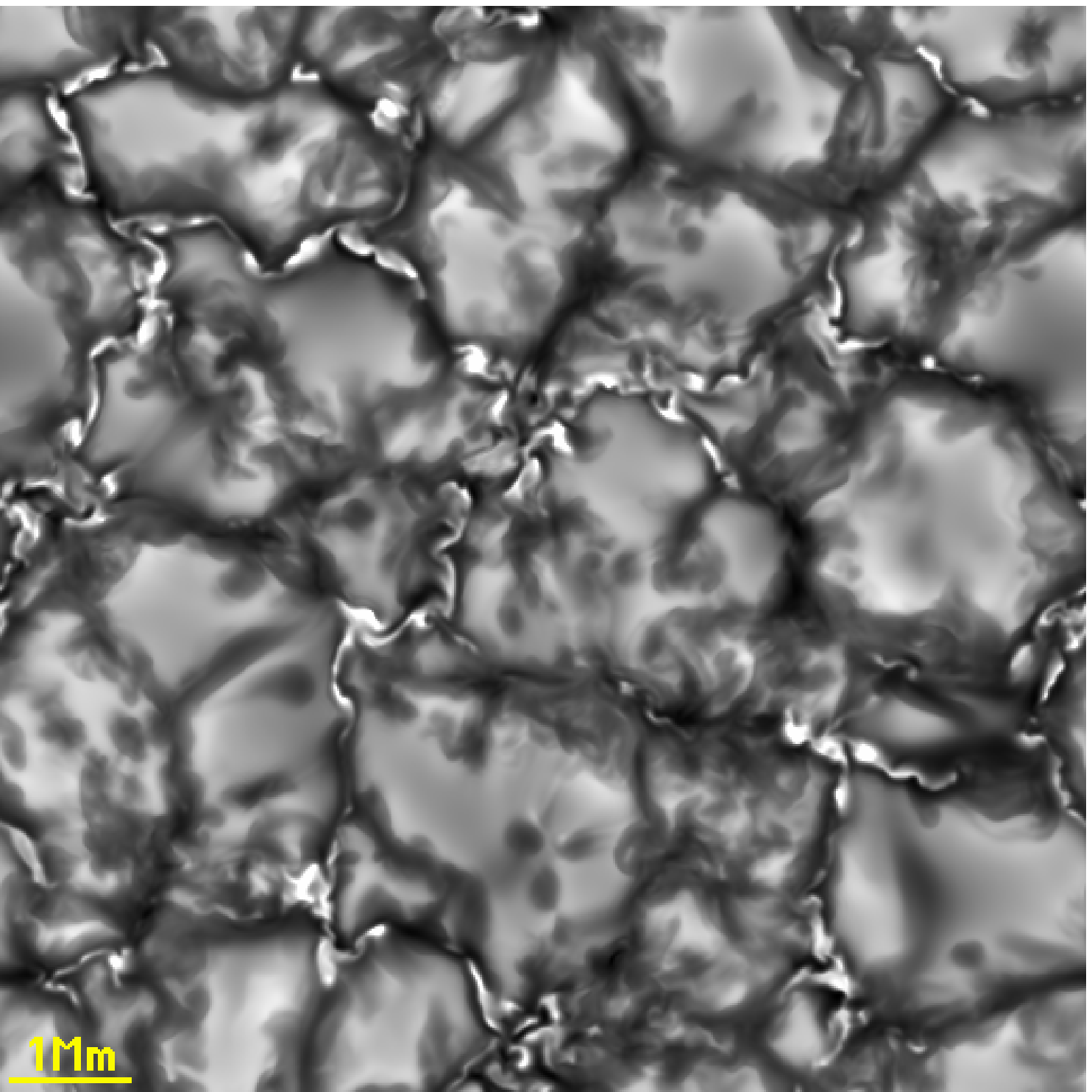} & \includegraphics[width=3.7cm]{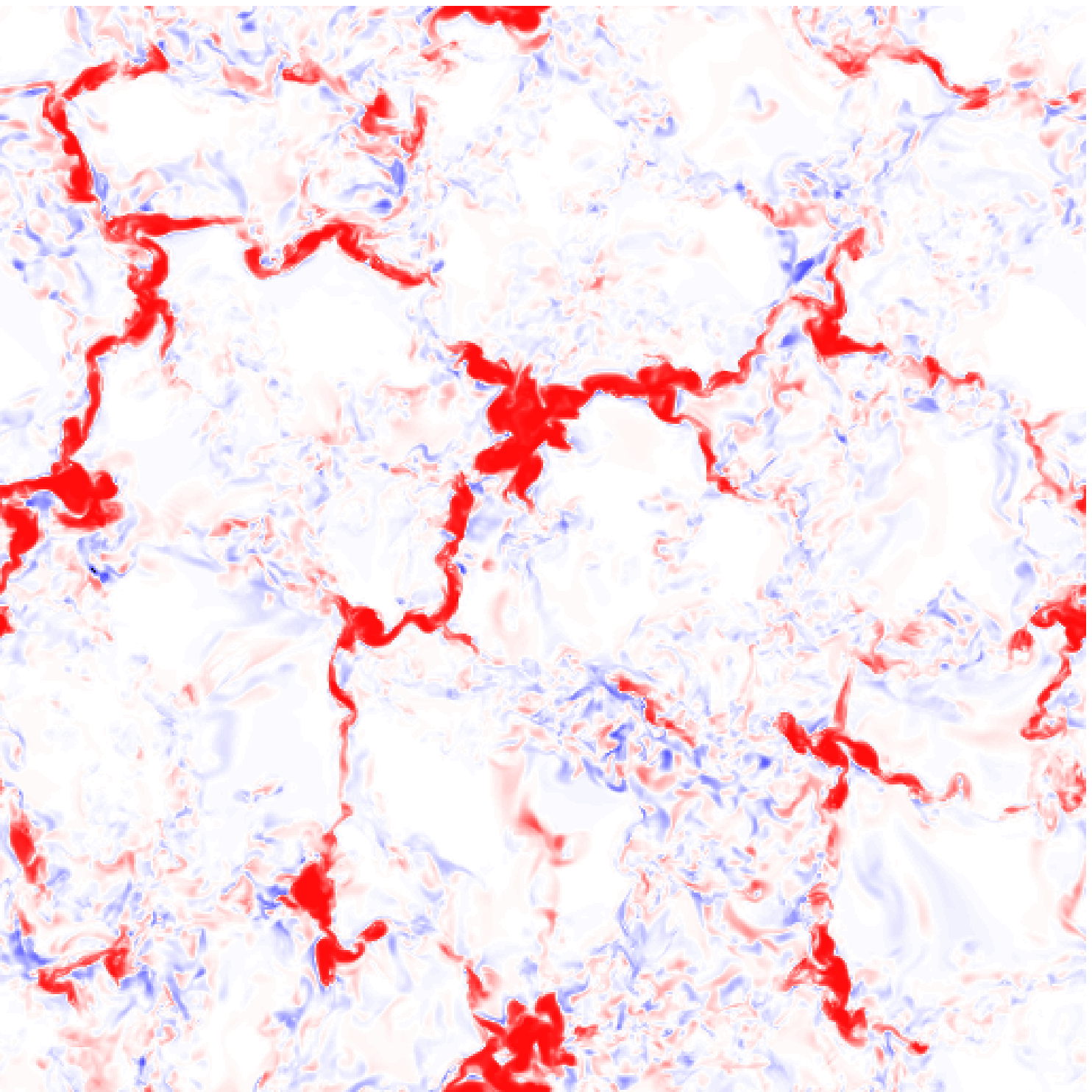}\\\hline
\multicolumn{2}{c}{$I_{\mathrm{bol}}(\mu=1)$ \qquad\qquad M2V \qquad\qquad $B_z(\tau_{\mathrm R}\approx 1)$}\\
\includegraphics[width=3.7cm]{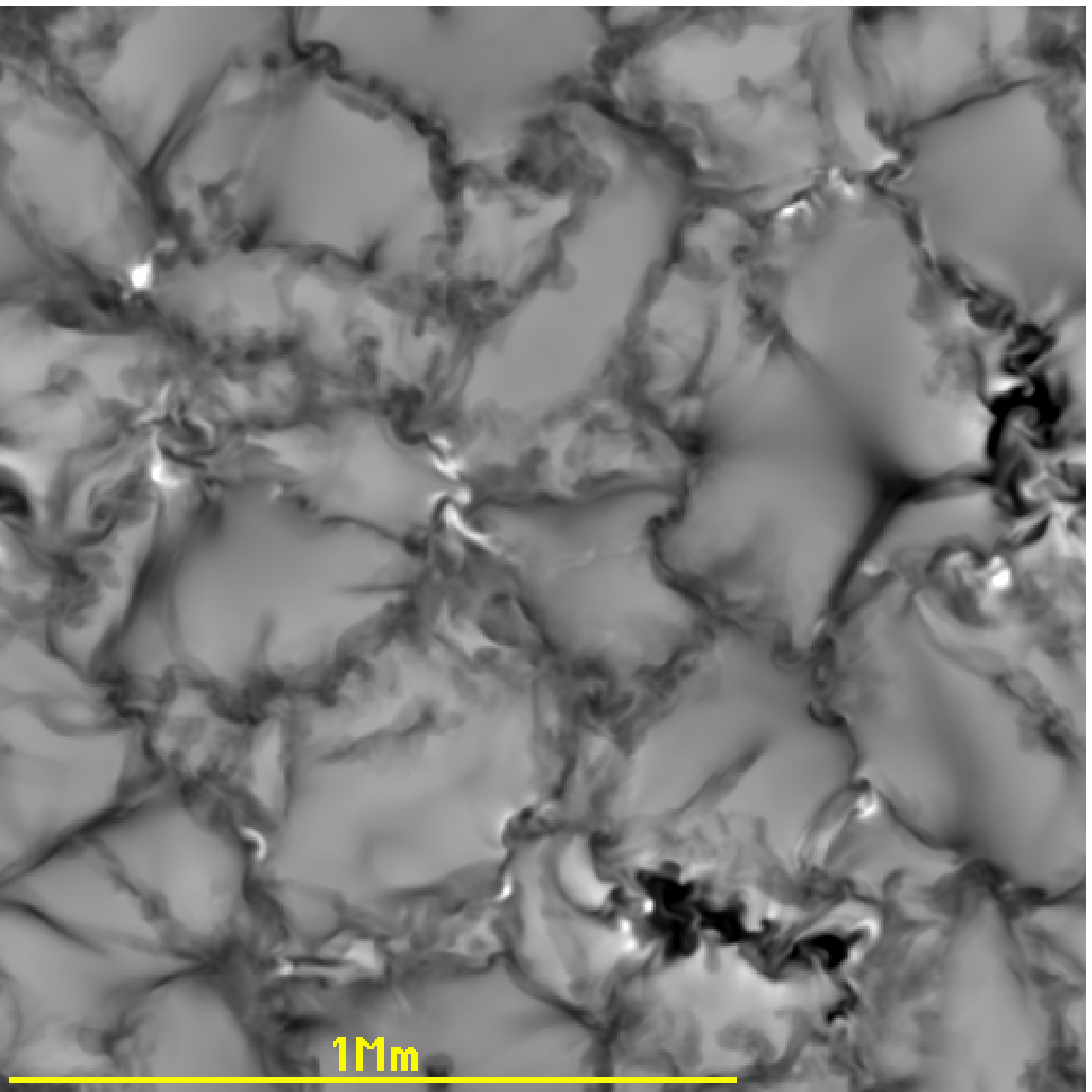} & \includegraphics[width=3.7cm]{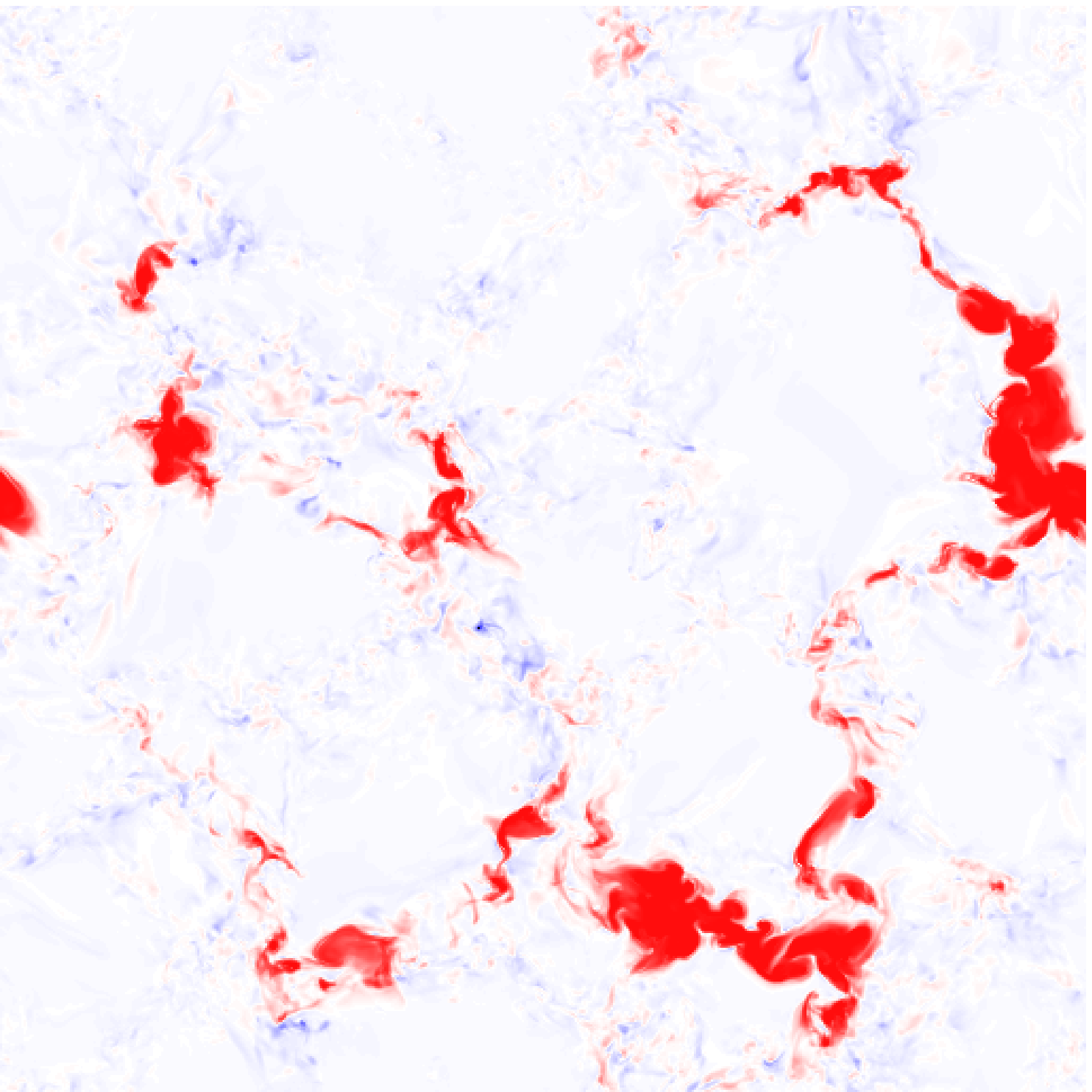}
\end{tabular}
\caption{ Snapshots of the bolometric intensity $I_{\mathrm{bol}}(\mu=1)$ (right panels) and vertical component $B_z(\tau_{\mathrm{R}}\approx 1)$ of the magnetic field at the optical surface (left panels) for MHD models of (G2V, upper panels) and an M2 dwarf (lower panels).}\label{fig:mag.comp}
\end{figure}
The modelled magneto-convection shows significant differences between
M-dwarfs and stars of earlier spectral types, e.\,g. the Sun. As
illustrated by Fig.~\ref{fig:mag.comp}, the initially homogeneous
magnetic flux is accumulated into very few structures of high field
strength, the cause of which are stable downflows.
While solar magnetic structures appear as bright features,
the magnetic structures on M-dwarfs tend to be rather dark. In the
case of the Sun, magnetic structures create a strong depression of the
optical surface with hot side walls that can radiatively heat the
interior of the magnetic structure. Owing to higher densities and shallower temperature gradient, this side wall heating is much less
efficient for the magnetic structures in M-dwarf atmospheres. Since the magnetic field suppresses
convective energy transport, the structures cool down.\\ 
These findings indicate that plage regions on M-stars
might not show bright points but rather ``pores'' and small ``star
spots'' of reduced intensity, which has a crucial impact on the
interpretation of observational data such as M-dwarf spectra.

\section{Generalized 3-D Radiative Transfer for Astrophysical Atmospheres (E.~Baron)}\label{s:baron}

\texttt{PHOENIX} is a generalized model atmosphere code which works in
1 or 3 spatial dimensions. The philosophy behind \texttt{PHOENIX} is
that it should work in a wide range of astrophysical environments and
that it should handle both static and moving flows in full
relativity. \texttt{PHOENIX} is
well calibrated on many astrophysical objects: Planets/BDs, Cool
Stars, Hot Stars ($\beta$CMa, $\epsilon$CMa), $\alpha$-Lyrae, Novae,
and SNe (Iabc, IIP, IIb) (see Hauschildt \& Baron 2010 an references therein). 
%\citep[see][and references therein]{hb10a}.
Much of the development work on \texttt{PHOENIX} has been devoted to
handling  radiative transfer in velocity flows.   Velocities are important
in many astrophysical objects: novae,    supernovae,    AGN, and
$\gamma$-ray bursts. But of course velocities are also important in 
  stars since the linewidth is determined by the convective velocity
  field as shown by Stein \& Nordlund (2000).
  %\citet{SN00}. 

There are two ways to deal with velocity fields: the Eulerian
formulation and the co-moving formulation. Each has advantages and
disadvantages. In the Eulerian formulation wavelengths are uncoupled,
significantly reducing memory requirements; however, opacities are
angle dependent, significantly increasing computational
requirements. It is also extremely difficult to handle relativity in
the Eulerian formulation. In the co-moving formulation one can include
both special and general relativity exactly and opacities are
isotropic, significantly reducing computational requirements; however, 
wavelengths are coupled, significantly increasing memory requirements.
While the solution of the radiative transfer equation in the co-moving
frame in 1-D has been understood for quite some time (Mihalas 1980);
%\citep{mih80}; 
in 3-D it is much  more complex and has been mostly approached via the
cumbersome tetrad formalism (Morita \& Kaneko 1984, 1986).
%\citep{morita84,morita86}.  
A much simpler
approach via affine parameters was developed in (Chen et al. 2007)
%\citet{bin07} and
and implemented in (Baron et al. 2009).
%\citet{bhb09}. 
The Eulerian Formulation is valid for velocities $v <
1000$~km~s$^{-1}$ and thus is of interest in stars with low
velocities. The Eulerian formulation trades off the high memory
requirement of the co-moving frame for the explicit coupling of angles
and frequencies, that is, all momentum space variables are
coupled. Thus, at each spatial point opacities must be calculated for
each coupled wavelength direction point, leading to both a large
amount of computation and storage. Nevertheless \texttt{PHOENIX} has
been adapted to include the Eulerian formulation in 3-D (Seelmann et
al. 2010).
%\citep{shb10}.

In summary, 
  \texttt{PHOENIX 3-D} solves the generalized atmosphere problem with both
    co-moving and 
    Eulerian formulations in Cartesian, spherical, and cylindrical geometry.
We still need to study which approach is computationally better, which
    may depend on the particular computer architecture, particularly
    with the advent of GPUs and very low memory per core exoscale computing.
The next step is to go beyond test problems to production code. 
While full 3-D RT is too computationally complex for radiation
    hydrodynamics, some of the methods we have developed may be adapted to a
    more simplified approach.
It is crucial  to do full radiative transfer to determine abundances,
perform detailed hydrodynamic model verification, and for other
applications to observed data.

\section{Multi-D hydro-simulations of substellar atmospheres (Adam P.Showman)}\label{s:showman}

 Over 100 transiting hot Jupiters are now known, and
observations from the Spitzer and Hubble Space Telescopes and
groundbased facilities constrain the
atmospheric composition and three-dimensional temperature structure
of many such objects.  Phase curves show that some hot Jupiters,
such as HD 189733b, have modest ($\sim$200 K) day-night temperature 
variations (e.g., Knutson et al. 2007), while others have
much larger day-night temperature differences.  For the case
of HD 189733b --the best-observed hot Jupiter-- the Spitzer infrared 
light curves imply that the hottest region is not at the substellar point
but rather is displaced ~30 degrees of longitude to the east. 
This feature provides strong evidence of atmospheric circulation
on these tidally locked planets.  

The atmospheric dynamical regime of hot Jupiters differs from
that of, for example, brown dwarfs.  The atmospheric circulation
on hot Jupiters is probably driven primarily by the $\sim 10^5 - 10^6$ W/m$^2$
net radiative heating on the dayside and cooling on the nightside;
unlike the case of Jupiter or typical brown dwarfs, the absorbed
stellar flux exceeds the convective fluxes in the planet's interior by ~3-5
orders of magnitude.  Evolution and structure models indicate that
multi-Byr-old hot Jupiters have deep radiative zones extending
from the top of the atmosphere to pressures of typically ~100-1000 bars.
Thus, the weather near the infrared photosphere on hot Jupiters occurs 
in a stably stratified radiative zone.  Hot Jupiters are thought
to be synchronously rotating with their ~1-10-day orbital periods,
implying that planetary rotation is less dominant in their dynamics
than is the case for Jupiter or typical brown dwarfs.

A variety of 3D dynamical models of the atmospheric circulation on hot
Jupiters have been published (e.g., Showman and Guillot 2002;
Dobbs-Dixon \& Lin 2008; Showman et al. 2008, 2009; Rauscher \& Menou
2010; Thrastarson and Cho 2010).  These models typically model the
circulation of hot Jupiters on 2-4-day orbits assuming the interior is
tidally locked.  At the pressure of the infrared photosphere, the
circulation in these models typically exhibits a banded structure,
with ~1-3 broad east-west jet streams whose speeds reach several
km/sec.  Day-night temperature differences are commonly hundreds of K
at the photosphere.  Real hot Jupiters probably exhibit a wide
diversity of behaviors, which have yet to be thoroughly explored in
circulation models.
%, and the parameter space has not yet been well explored in these models.

\begin{figure}[!ht]
\centerline{\includegraphics[width=8cm]{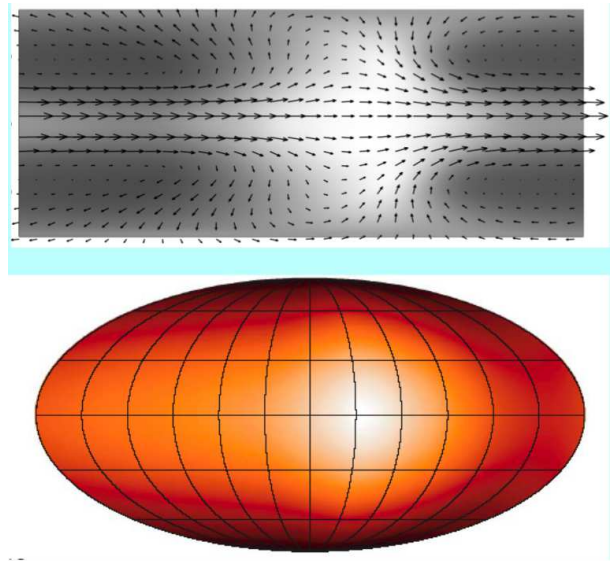}}
%\plotone{agol/cs16_agol_figure.eps}
%\plotfiddle{CS16_HD189733.pptx.ps}{5in}{0.}
\caption{ (Bottom): Longitudinal temperature structure of
of hot Jupiter HD 189733b inferred from Spitzer 8-micron
infrared light curve (Knutson et al. 2007), showing eastward
offset of hottest region from substellar longitude.  (Top)
Temperature pattern (greyscale) and winds (arrows) from
three-dimensional circulation model of a hot Jupiter by
Showman and Guillot (2002).  A common feature of such models
is the eastward equatorial jet, which displaces the hottest
regions to the east of the substellar point, as seen in the
observations.}
\label{fig:showman}
\end{figure}

Interestingly, the eastward offset of the hot region from the 
substellar longitude inferred from infrared light curves of HD 189733b 
-- which provides our current best evidence for atmospheric circulation on
hot Jupiters -- was predicted five years before its discovery
(Showman and Guillot 2002).  In their model, the eastward offset
results from advection by a robust, eastward flowing equatorial jet stream 
that dominates the circulation (Fig.~\ref{fig:showman}).  Subsequent models by several groups
have generally confirmed the robustness of this feature 
%for the parameter regime of typical hot Jupiters 
(e.g., Showman et al. 2008,
2009; Dobbs-Dixon and Lin 2008; Rauscher
and Menou 2010).  However, to date, the mechanism for this so-called
"equatorial superrotation" has not been identified.  New, unpublished
work by A.P. Showman and L.M. Polvani shows, however, that the
superrotation results from standing Rossby waves generated by
the day-night thermal forcing.  The Rossby waves, which are planetary
in scale, generate phase tilts such that equatorward-moving air
exhibits greater eddy angular momentum than poleward-moving air.
This pumps eddy angular momentum from the midlatitudes to the equator
and generates the equatorial superrotating jet.  Showman and Polvani
demonstrated the mechanism in an idealized, linear, analytic
model, in simplified nonlinear models, and in full three-dimensional
general circulation models.  An implication is that the mechanism
for producing the jet need not involve turbulent cascades or other 
eddy-eddy interactions, but rather results from a direct interaction 
between the standing, thermally generated quasi-linear eddies 
and the mean flow at the planetary scale.

\section{ Weather on a Hot Jupiter (Eric Agol)}\label{s:agol}

Testing global models of weather on hot jupiters requires
observations of their global properties.  To date, the 
best means available for such a comparison is infrared observations
of the phase variation of hot jupiters (Knutson et al. 2007)\nocite{Knutson2007}.
If the weather pattern
changes on a timescale slower than the orbital time of the
planet, then during the orbit the different faces of the planet
will be observed, allowing a deconvolution of the longitudinal
brightness of the planet (Cowan \& Agol 2008)\nocite{Cowan2008}.  Since the planet
cannot be resolved from the star, there are several possible ways 
such an analysis might go wrong: (1)
stellar variability might swamp the planet phase variation;
(2) planet variability might invalidate the steady-state assumption
required for inversion;  (3) planet-star interaction might
cause stellar brightness variations on a similar timescale
as the planet's orbit.  In addition, instrumental effects
can be present which may be stronger than the phase variation.
So far the best target for this sort of observation is the
exoplanet HD 189733b.  It orbits a bright star, it is large
in size compared to its host star, and it is hot enough and
has a short enough period to enable observations of a
reasonable duration.  However, the host star is strongly
variable in the optical, $\sim 1-2$\%, with a period of
about 12 days, so the host star variability must be accounted
for to properly measure the planet's infrared variability
with the Spitzer Space Telescope.  We measured the phase
variation over slightly longer than half an orbital
period at 8 $\mu$m with IRAC Channel 4 (Knutson et al. 2007),
and then observed a subsequent six transits and six eclipses
with the goals of determining the day-side variability
and looking for transit-timing variations (Agol et al. 2008,
Agol et al. 2010)\nocite{Agol2009, Agol2010, Cowan2010}.  We also obtained simultaneous ground-based
monitoring in the optical which we used to correct for
the stellar variability by extrapolating the optical stellar
variation into the infrared (Winn \& Henry 2008)\nocite{Henry2008}.

\begin{figure}[!ht]
\centerline{\includegraphics[width=8cm]{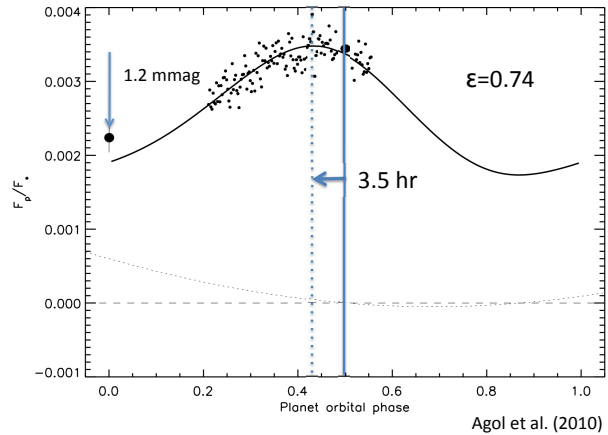}}
%\plotone{agol/cs16_agol_figure.eps}
%\plotfiddle{CS16_HD189733.pptx.ps}{5in}{0.}
\caption{Measured phase-variation of HD 189733b (dots)
at 8 $\mu$m after correction for stellar variability (dotted 
line). The peak of the phase function is offset 3.5 hours
before secondary eclipse (orbital period: 53 hours).
The night-side is 1.2 mmag fainter than the dayside, or
about 64\% of the flux.  The phase variation may be
fit by a toy model (solid line; Cowan \& Agol 2010) in which 
the energy is advected in a super-rotating jet in which
the ratio of the radiative to advection times $\epsilon = 0.74$.}
\label{fig:agol}
\end{figure}

Based on these data, we found that the absolute flux of
the system could be measured to $<$0.35 mmag after decorrelating
with instrumental variations and stellar variability.  We 
used this decorrelation to correct for the stellar variability,
giving a more precise phase variation (Figure~\ref{fig:agol}).  The observed 
phase function is in good qualitative agreement with models
of weather on this hot jupiter (e.g. Showman et al. 2009)\nocite{Showman2009},
albeit with an observed peak of the planet's flux that is closer to
the secondary eclipse than predicted by the models.  The
location of this peak is primarily controlled by the ratio
of the radiative timescale to the advection timescale, $\epsilon$, so
the data indicate that the models either overpredict the 
super-rotation speed of the equatorial jet, or they overpredict 
the cooling timescale at the 8 micron photosphere.  We
also find an offset in the secondary eclipse time; after correcting
for light-travel time across the system, the offset can be
accounted for by the asymmetric dayside flux caused by
the super-rotating jet (this offset is due to the fact that
we fit the secondary eclipse with a model in which the planet
is uniform in surface brightness).  The night side is about
64\% of the brightness of the day side and the secondary
eclipse depth variation has an RMS of $<$ 2.7\%, which is
limited by the photometric precision of the data.  These
results are in good agreement with general circulation models 
for this planet which predict fluctuations of $<$1\%.

\section{Overshoot, gravity waves and
  non-equilibrium chemistry (Derek Homeier, France Allard, Bernd Freytag)}\label{s:homeier}

The \texttt{PHOENIX} BT-Settl models (Allard et al. 2010)
%\citep{franceCS16} 
combine a 
cloud formation and gas phase non-equilibrium chemistry model, 
using vertical diffusion
profiles based on CO5BOLD RHD simulations as an input to our models,
finding the mixing in the transition zone from carbon monoxide- to
methane-dominated chemistry to be governed by gravity waves forming in
the upper atmosphere (Freytag et al. 2010).
%\citep{bdCO5BOLD,berndCS16}. 
We introduce updated reaction rates for the CO to CH$_4$ conversion
from Visscher et al. (2010)
%\citet{visscherJup2010} 
and include departures from chemical
equilibrium in the CO to CO$_2$ ratio to determine the molecular
fractions of CH$_4$, CO and CO$_2$ for atmospheres spanning the range
from L to T dwarfs. Synthetic spectra calculated for the resulting
compositions reproduce the observed upmixing of carbon monoxide in
brown dwarfs across the L/T transition. In addition we find carbon
dioxide to appear in excess of its CE abundance in T dwarfs. The
models produce an improved fit to the observed mid-infrared photometry
of the coolest brown dwarfs (Burningham et al. 2010),
%\citep{benRoss458C}, 
and are confirmed by
the identification of distinctive 4.2\,$\mu$m CO$_2$ absorption features in
several late T dwarf spectra by the AKARI satellite  (Yamamura et al. 2010).
%\citep{akariCO2}. 

\acknowledgements 
{\small EB was supported by
NSF grant AST-0707704,  US DOE Grant
DE-FG02-07ER41517, and by  
program number HST-GO-12298.05-A which is supported by NASA through 
a grant from the Space Telescope Science Institute, which is operated by the 
Association of Universities for Research in Astronomy, Incorporated, under 
NASA contract NAS5-26555. This research used resources of the NERSC, which is supported by the Office
of Science of the U.S.  Department of Energy under Contract No.
DE-AC02-05CH11231; and the H\"ochstleistungs Rechenzentrum Nord (HLRN).
Support for EA was provided by NSF through
CAREER Grant No. 0645416 and by NASA through an
award issued by JPL/Caltech.  APS acknowledges funding from NASA
Origins.} D.~H.\ gratefully acknowledges support from a foreign travel
grant by the DAAD. F.~A.\ and
B.~F.\ acknowledge financial support from the ANR, and the
PNPS of CNRS/INSU, France. Stellar and substellar
atmosphere models for this study have been calculated at the GWDG.

%\bibliography{Helling_Ch}

\begin{thebibliography}{}
\expandafter\ifx\csname natexlab\endcsname\relax\def\natexlab#1{#1}\fi
\expandafter\ifx\csname url\endcsname\relax
  \def\url#1{\texttt{#1}}\fi
\expandafter\ifx\csname urlprefix\endcsname\relax\def\urlprefix{URL }\fi
\providecommand{\eprint}[2][]{\url{#2}}

\bibitem[{{Agol} et~al.(2009){Agol}, {Cowan}, {Bushong}, {Knutson},
  {Charbonneau}, {Deming}, \& {Steffen}}]{Agol2009}
{Agol}, E., {Cowan}, N.~B., {Bushong} et al. 2009, vol. 253 of IAU
  Symposium, 209

\bibitem[{{Agol} et~al.(2010){Agol}, {Cowan}, {Knutson}, {Deming}, {Steffen},
  {Henry}, \& {Charbonneau}}]{Agol2010}
{Agol}, E., {Cowan}, N.~B., {Knutson}, H.~A. et al. D. 2010, \apj, 721, 1861

\bibitem[{{Allard} et~al.(2010){Allard}, {Homeier}, \& {Freytag}}]{franceCS16}
{Allard}, F., {Homeier}, D., \& {Freytag}, B. 2010 {\em this vol.}\eprint{1011.5405} 

\bibitem[{Baron et~al.(2009) Baron, Hauschildt, \& Chen}]{bhb09}
Baron, E., Hauschildt, P.~H., \& Chen, B. 2009, A\&A, 498, 987

\bibitem[]{} Berdyugina S.~V., Berdyugin A.~V., Fluri D.~M., Piirola V. 2011, ApJ submitted

\bibitem[ ]{} Berdyugina S.~V., in Polarimetry of cool atmospheres: From the Sun to exoplanets, eds. J.R.~Kuhn, 2011, (arXiv:1011.0751)

\bibitem[ ] {}{Berdyugina}, S.~V.,  {Berdyugin}, A.~V.,  {Fluri}, D.~M., {Piirola}, V. 2008, APJ 673, L83

\bibitem[]{} {Berdyugina}, S.~V. 2005, Liv. Rev. Solar Phys. 2, 8

\bibitem[{{Berger} {et~al.}(2001){Berger}, {Haguenauer}, {Kern}, {Perraut},
  {Malbet}, {Schanen}, {Severi}, {Millan-Gabet}, \&
  {Traub}}]{2001A&A...376L..31B}
{Berger}, J.~P., {Haguenauer}, P., {Kern}, P. et al. 2001, \aap,
  376, L31
  
 \bibitem[{{Burningham} et~al.(2010){Burningham}, {Leggett}, {Homeier},
  {Saumon}, {Lucas}, {Pinfield}, {Tinney}, {Allard}, Marley, {Jones}, Murray,
  {Ishii}, {Day-Jones}, Gomes, \& Zhang}]{benRoss458C}
{Burningham}, B., {Leggett}, S.~K., {Homeier} et al. 2010, submitted to \mnras.
  
\bibitem[{{Chelli} {et~al.}(2009){Chelli}, {Duvert}, {Malbet}, \&
  {Kern}}]{2009A&A...498..321C}
{Chelli}, A., {Duvert}, G., {Malbet}, F., \& {Kern}, P. 2009, \aap, 498, 321

\bibitem[{Chen et~al.(2007)Chen, Kantowski, Baron, Knop, \& Hauschildt}]{bin07}
Chen, B., Kantowski, R., Baron, E., Knop, S., \& Hauschildt, P. 2007, MNRAS,
  380, 104

\bibitem[{{Cowan} \& {Agol}(2008)}]{Cowan2008}
{Cowan}, N.~B., \& {Agol}, E. 2008, \apjl, 678, L129

\bibitem[{{Cowan} \& {Agol}(2010)}]{Cowan2010}
--- 2010, arXiv:1011.0428

\bibitem[] {}Dobbs-Dixon, I. \& Lin, D.N.C. 2008. ApJ 673, 513.

%\bibitem[{{Deleuil} {et~al.}(2009){Deleuil}, {Deeg}, {Alonso}, {Bouchy},
%  {Rouan}, {Auvergne}, {Baglin}, {Aigrain}, {Almenara}, {Barbieri}, {Barge},
%  {Bruntt}, {Borde}, {Collier Cameron}, {Csizmadia}, {de La}, {Dvorak},
%  {Erikson}, {Fridlund}, {Gandolfi}, {Gillon}, {Guenther}, {Guillot}, {Hatzes},
%  {Hebrard}, {Jorda}, {Lammer}, {Leger}, {Llebaria}, {Loeillet}, {Mayor},
%  {Mazeh}, {Moutou}, {Ollivier}, {Paetzold}, {Pont}, {Queloz}, {Rauer},
%  {Schneider}, {Shporer}, {Wuchterl}, \& {Zucker}}]{2009yCat..34910889D}
%{Deleuil}, M., {Deeg}, H.~J., {Alonso}, R. et al. 2009, VizieR Online Data
%  Catalog, 349, 10889

\bibitem[{{Fossati} et~al.(2010{\natexlab{a}}){Fossati}, { Haswell}, {Froning},
  \& {et~al.}}]{fossati2010}
{Fossati}, L., { Haswell}, C.~A., {Froning}, C.~S., \& {et~al.}
  2010{\natexlab{a}}, \apjl, 714, L222

\bibitem[{{Fossati} et~al.(2010{\natexlab{b}}){Fossati}, {Bagnulo}, {Elmasli},
  \& {et~al.}}]{fossati2010b}
{Fossati}, L., {Bagnulo}, S., {Elmasli}, A., \& {et~al.} 2010{\natexlab{b}},
  \apj, 720, 872

\bibitem[{{Freytag} et~al.(2010){Freytag}, {Allard}, {Ludwig}, {Homeier}, \&
  {Steffen}}]{bdCO5BOLD}
{Freytag}, B., {Allard}, F., {Ludwig}, H., {Homeier}, D., \& {Steffen}, M.
  2010, \aap, 513
  
%\bibitem[{{Freytag} et~al.(2010b){Freytag}, {Allard}, {Ludwig}, {Homeier}, \&
%  {Steffen}}]{berndCS16}
%{Freytag}, B., {Allard}, F., {Ludwig}, H., {Homeier}, D., \& {Steffen}, M.
%  2010b, {\em this vol.\ suppl.}


\bibitem[{Hauschildt \& Baron(2010)}]{hb10a}
Hauschildt, P.~H., \& Baron, E. 2010, A\&A, 509, A36

\bibitem[{{Hebb} et~al.(2009){Hebb}, {Collier-Cameron}, {Loeillet}, \&
  {et~al.}}]{hebb2009}
{Hebb}, L., {Collier-Cameron}, A., {Loeillet}, B., \& {et~al.} 2009, \apj, 693,
  1920

\bibitem[{{Henry} \& {Winn}(2008)}]{Henry2008}
{Henry}, G.~W., \& {Winn}, J.~N. 2008, \aj, 135, 68

\bibitem[{{Homeier} et~al.(2010){Homeier}, {Freytag} \& {Allard}}]{derekCS16}
{Homeier}, D., {Freytag}, B., \& {Allard}, F. 2010 {\em this vol.\ suppl.}

\bibitem[{{Kloppenborg} {et~al.}(2010){Kloppenborg}, {Stencel}, {Monnier},
  {Schaefer}, {Zhao}, {Baron}, {McAlister}, {Ten Brummelaar}, {Che},
  {Farrington}, {Pedretti}, {Sallave-Goldfinger}, {Sturmann}, {Sturmann},
  {Thureau}, {Turner}, \& {Carroll}}]{2010Natur.464..870K}
{Kloppenborg}, B., {Stencel}, R., {Monnier}, J.~D. et al. 2010, \nat, 464, 870


\bibitem[{{Knutson} et~al.(2007){Knutson}, {Charbonneau}, {Allen}, {Fortney},
  {Agol}, {Cowan}, {Showman}, {Cooper}, \& {Megeath}}]{Knutson2007}
{Knutson}, H.~A., {Charbonneau}, D., {Allen}, L.~E. et al.  2007, \nat, 447, 183


\bibitem[{{Kraus} {et~al.}(2010){Kraus}, {Hofmann}, {Menten}, {Schertl},
  {Weigelt}, {Wyrowski}, {Meilland}, {Perraut}, {Petrov}, {Robbe-Dubois},
  {Schilke}, \& {Testi}}]{2010Natur.466..339K}
{Kraus}, S., {Hofmann}, K., {Menten}, K.~M. et al. 2010, \nat, 466, 339

\bibitem[] {} {Kuhn}, J.~R., {Geiss}, B.,  {Harrington}, D.~M. 2011,  in Solar Polarization 6, eds. Kuhn et al., (arXiv:1010.0705)

\bibitem[] {}{Kuhn}, J.~R., {Berdyugina}, S.~V., {Fluri}, D.~M., 
        {Harrington}, D.~M.,  {Stenflo}, J.~O. 2007, ApJL 668, L63


\bibitem[{{Lai} et~al.(2010){Lai}, {Helling}, \& {van den Heuvel}}]{lai2010}
{Lai}, D., {Helling}, Ch., \& {van den Heuvel}, E.~P.~J. 2010, \apj, 721, 923

%\bibitem[] {} Menou, K. \& Rauscher, E. 2009, ApJ 700, 887

\bibitem[{Mihalas(1980)}]{mih80}
Mihalas, D. 1980, ApJ, 237, 574

\bibitem[{{Monnier} {et~al.}(2003){Monnier}, {Berger}, {Millan-Gabet}, {Traub},
  {Carleton}, {Pedretti}, {Coldwell}, \& {Papaliolios}}]{2002SPIE.4838..1127}
{Monnier}, J.~D., {Berger}, J.~P., {Millan-Gabet}, R.  2003, in Proc. SPIE, Interferometry for Optical Astronomy, W.A. Traub,
  editor, Vol. 4838, 1127--1138

\bibitem[{{Monnier} {et~al.}(2007){Monnier}, {Zhao}, {Pedretti}, {Thureau},
  {Ireland}, {Muirhead}, {Berger}, {Millan-Gabet}, {Van Belle}, {ten
  Brummelaar}, {McAlister}, {Ridgway}, {Turner}, {Sturmann}, {Sturmann}, \&
  {Berger}}]{2007Sci...317..342M}
{Monnier}, J.~D., {Zhao}, M., {Pedretti}, E. et al. 2007, Science, 317, 342


\bibitem[{Morita \& Kaneko(1984)}]{morita84}
Morita, K., \& Kaneko, N. 1984, Ap\&SS, 107, 333

\bibitem[{Morita \& Kaneko(1986)}]{morita86}
--- 1986, Ap\&SS, 121, 105

\bibitem[{{Pedretti} {et~al.}(2009{\natexlab{a}}){Pedretti}, {Monnier},
  {Brummelaar}, \& {Thureau}}]{2009NewAR..53..353P}
{Pedretti}, E., {Monnier}, J.~D., {Brummelaar}, T.~T., \& {Thureau}, N.~D.
  2009{\natexlab{a}}, \nat, 53, 353

\bibitem[{{Pedretti} {et~al.}(2009{\natexlab{b}}){Pedretti}, {Monnier},
  {Lacour}, {Traub}, {Danchi}, {Tuthill}, {Thureau}, {Millan-Gabet}, {Berger},
  {Lacasse}, {Schuller}, {Schloerb}, \& {Carleton}}]{2009MNRAS.397..325P}
{Pedretti}, E., {Monnier}, J.~D., {Lacour}, S. et al. 2009{\natexlab{b}}, \mnras, 397, 325

%\bibitem[{{Ragland} {et~al.}(2008){Ragland}, {Le Coroller}, {Pluzhnik},
%  {Cotton}, {Danchi}, {Monnier}, {Traub}, {Willson}, {Berger}, \&
%  {Lacasse}}]{2008ApJ...679..746R}
%{Ragland}, S., {Le Coroller}, H., {Pluzhnik}, E. et al. 2008, \apj, 679, 746

\bibitem[] {} Rauscher, E. \& Menou, K. 2010, ApJ 714, 1334


\bibitem[{Seelmann et~al.(2010)Seelmann, Hauschildt, \& Baron}]{shb10}
Seelmann, A., Hauschildt, P.~H., \& Baron, E. 2010, A\&A, 522, A102


\bibitem[{{Scholz} {et~al.}(2003){Scholz}, {McCaughrean}, {Lodieu}, \&
  {Kuhlbrodt}}]{2003A&A...398L..29S}
{Scholz}, R., {McCaughrean}, M.~J., {Lodieu}, N., \& {Kuhlbrodt}, B. 2003,
  \aap, 398, L29

\bibitem[{{Showman} et~al.(2009){Showman}, {Fortney}, {Lian}, {Marley},
  {Freedman}, {Knutson}, \& {Charbonneau}}]{Showman2009}
{Showman}, A.~P., {Fortney}, J.~J., {Lian}, Y. et al. 2009, \apj, 699, 564

\bibitem[] {}Showman, A.P. et al. 2008, ApJ 682, 559

\bibitem[] {}Showman, A.P., \& Guillot T. 2002, A\&A 385, 166.

\bibitem[{Stein \& Nordlund(2000)}]{SN00}
Stein, R.~F., \& Nordlund, {\r{A}}. 2000, Sol.~Phys., 192, 91

\bibitem[] {}Thrastarson, H.T. \& Cho, J.Y-K. 2010 ApJ, 716, 144


\bibitem[{{Vidotto} et~al.(2010{\natexlab{a}}){Vidotto}, {Jardine}, \&
  {Helling}}]{paper1}
{Vidotto}, A.~A., {Jardine}, M., \& {Helling}, Ch. 2010{\natexlab{a}}, \apjl,
  722, L168 (Paper 1)

\bibitem[{{Vidotto} et~al.(2010{\natexlab{b}}){Vidotto}, {Jardine}, \&
  {Helling}}]{paper2}
--- 2010{\natexlab{b}}, MNRAS Letter, in press. \eprint{arXiv: 1011.3455}

\bibitem[{{Visscher} et~al.(2010){Visscher}, {Moses}, \&
  {Saslow}}]{visscherJup2010}
{Visscher}, C., {Moses}, J.~I., \& {Saslow}, S.~A. 2010, Icarus, 209, 602.
  \eprint{1003.6077}
  
\bibitem[{{V{\"o}gler} et~al.(2005){V{\"o}gler}, {Shelyag}, {Sch{\"u}ssler},
  {Cattaneo}, {Emonet}, \& {Linde}}]{MURaM1}
{V{\"o}gler}, A., {Shelyag}, S., {Sch{\"u}ssler}, M. et al. 2005, \aap, 429, 335

\bibitem[] {}{Wiktorowicz}, S.~J. 2009, ApJ 696, 1116

\bibitem[] {}Woitke, P. \& Helling, Ch 2004, A\&A 414, 335

\bibitem[{{Yamamura} et~al.(2010){Yamamura}, {Tsuji}, \&
  {Tanab{\'e}}}]{akariCO2}
{Yamamura}, I., {Tsuji}, T., \& {Tanab{\'e}}, T. 2010, \apj, 722, 682.
  \eprint{1008.3732}

\bibitem[{{Zhao}(2009)}]{2009PhDT.........7Z}
{Zhao}, M. 2009, PhD thesis, University of Michigan

\end{thebibliography}

\end{document}